\documentclass[12pt]{iopart}
\usepackage{longtable}
\usepackage{graphicx,amssymb}

\begin{document}

\title[Few-electron semiconductor quantum dots with Gaussian confinement]{Few-electron semiconductor quantum dots with Gaussian confinement}

\author{Sergio S.~Gomez and Rodolfo H.~Romero}

\address{Facultad de Ciencias Exactas, Universidad Nacional del Nordeste, Avenida Libertad 5500 (3400) Corrientes, Argentina}
\ead{ssgomez@exa.unne.edu.ar, rhromero@exa.unn.edu.ar}
\begin{abstract}
We have performed Hartree-Fock calculations of electronic structure of $N\le 10$ electrons in a quantum dot modeled with a confining Gaussian potential well. We discuss the conditions for the stability of $N$ bound electrons in the system. We show that the most relevant parameter determining the number of bound electrons is $V_0R^2$. Such a property arises from widely valid scaling properties of the confining potential. Gaussian Quantum dots having $N=2$, 5 and 8 electrons are particularly stable in agreement of Hund rule. The shell structure becomes less and less noticeable as the well radius increases.
\end{abstract}

\submitto{\JPCM}
%\maketitle
%###################################################################################
\section{Introduction}
Modern semiconductor technology has allowed to fabricate and manipulate electrons confined within regions of nanometer size having a plethora of shapes. 
The possibility of tunning the shape, size and number of bound electrons of those nanostructures has raised a great deal of interest on the subject of the confined low-dimensional few-electron systems known as quantum dots (QDs) \cite{Harrison, Michler} either for specific applications or exploring new fundamental phenomena at a quantum level \cite{Hanson07, Awschalom07,Atature06}.

Quantum dots have been also termed ``artificial atoms'' because their electronic structure and properties resemble those of natural atoms \cite{Reimann02}. Electrons in quantum dots are confined due to potential barriers, in much the same way as electrons in atoms are confined due to Coulomb nuclear attraction. Further similarities arise because electrons within quantum dots interact with each other through Coulomb forces, their energy spectra present both discrete and continuum states giving rise to binding and dissociation processes, while transitions between them give rise to emission and absorption of radiation. 
Thus it is justified, to some extent, to apply --{\em mutatis mutandis}-- methods of atomic or molecular physics to the study of these systems of electrons acted upon by a given confining potential. A spherically symmetric confining potential can provide a model for semiconductor spherical nanocrystals embbeded within an insulator \cite{Alivisatos96}.
In the literature dealing with calculations of electronic structure and properties of QDs, a confining harmonic potential model is the most ubiquitous one. This can be justified on the basis of a generalized Kohn's theorem \cite{Kohn61} satisfied by such a potential and, approximately, verified by experiments. A potential of this type allows to explain the electronic shell structure observed, {\em e.g.}, in small clusters. Nevertheless, the infinite height and range of a parabolic potential is clearly unphysical and allows to accomodate an infinite number of bound electrons, what precludes the consideration of binding and dissociation processes. 
Other finite-range models have been also advocated; {\em e.g.}, 
the energy spectra of two- and three-electron systems in a spherical potential well of finite depth was obtained variationally \cite{Szafran99}, and the unrestricted Hartree-Fock (UHF) method was applied to the same model for the calculation of the electronic structure of systems having up to 20 electrons \cite{Bednarek99}. More recently, the UHF calculation of ground state, chemical potential and charging energies of electrons in an infinite spherical potential well, with and without magnetic field, was reported \cite{Destefani04}.
However, the sharp discontinuity at the QD radius for such a potential well is not completely satisfactory from neither a physical nor a computational point of view, and some interpolating potentials \cite{Ranjan02} as well as smoothly varying potentials have been proposed \cite{DeFilippo00, Adamowski00, Xie03, Boyacioglu07,Ciftja07}. Among them, the Gaussian potential has received some attention and its one- \cite{Bessis82, Lai83, Cohen84, Chatterjee85} and two-electron spectra \cite{Adamowski00, Xie03, Boyacioglu07} have been calculated. Furthermore, this potential is particularly suitable for an atomic-like treatment such as UHF calculations. 
The Hartree-Fock method is one of the most widely used methods for the calculation of electronic structure of atoms and molecules. Its usual implementation consists in the expansion of the unknown orbitals as a linear combination in a given basis set. The method thus determine self-consistently the energy eigenvalues and the orbitals. In atomic problems, the most common basis sets are spherical or Cartesian Gaussian functions because they provide closed expressions for the matrix elements of the atomic Hamiltonian. Such an advantage holds even more when the confining potential is Gaussian itself. Hence, we take in this work an atomic-like approach for the calculation of the electronic structure of few-electron Gaussian quantum dots. Such an approach shall be particularly advantageous when dealing with, {\em e.g.}, systems of coupled quantum dots, and also for the analisys of mixed systems, like molecules coupled to QDs. Work along that line will be published elsewhere. On the other hand, such a methodology is the starting point for a systematic improvement in the treatment of correlation effects by, {\em e.g.} higher-order perturbation theory or multi-reference methods \cite{Fulde, Szabo}.

\section{Theory} 
 
We consider a system of few electrons confined by a potential assumed in the form of a spherically symmetric Gaussian potential well of a typical radius $R$ and finite depth $-V_0$, {\em i.e.}, $V(r)=-V_0 \exp{(-r^2/2R^2)}$. 
It approaches a parabolic behavior around its minimum while goes smoothly to zero at infinity.
In Ref \cite{Adamowski00} the parabolic approximation 
\begin{equation}\label{harmonic_ap}
V(r) = -V_0 e^{-r^2/2R^2} \simeq -V_0 + V_0\frac{r^2}{2R^2}
  = -V_0 + \frac{1}{2}m\omega^2 r^2,
\end{equation}
was considered for comparison, where 
\begin{equation}
\label{omega}
\omega^2 = \frac{V_0}{mR^2} = \frac{2\lambda V_0}{m},
\end{equation}
relates the frequency $\omega$ of the harmonic oscillator corresponding to the Gaussian exponent $\lambda\equiv 1/2R^2$.
The Schr\"odinger equation for a system of $N$ electrons confined within the Gaussian well and interacting through Coulomb potentials is 
\begin{equation}
\label{sch eq}
\left[-\frac{\hbar^2}{2m^*}\sum_{i}\nabla_i^2+\sum_{i}V(r_i)+\sum_{i<j}\frac{e^2}{4\pi\kappa\epsilon_0 r_{ij}}\right]\psi =E\psi,
\end{equation}
where $m^*$ is the effective electron masses, and $\kappa$ is the medium dielectric constant $\kappa=\epsilon/\epsilon_0$. By introducing donor Bohr radius $a_D=(\kappa m/ m^*)a_B$ as the unit of length and donor Rydberg $R_D=(m^*/m\kappa^2)$ Ry as the unit of energy, Eq. (\ref{sch eq}) becomes
\begin{equation}
\label{sch eq 2}
\left[-\sum_{i}\nabla_i^2+\sum_{i}V(r_i)+\sum_{i<j}\frac{2}{r_{ij}}\right]\psi =\varepsilon\psi,
\end{equation}
such that the results can be easily transferred between different materials by properly changing $a_D$ and $R_D$; for instance, for a GaAs semiconductor QD, one has $a_D=6$ meV and $R_D=10$ nm.\\

The existence of bound states of Eq. (\ref{sch eq 2}) depends on the radius and depth of the Gaussian potential. If both magnitudes are small enough, even a single electron cannot form a stable bound state.
Some physical insight on the stability of one- and few-electron QDs can be gained from a simple variational estimate. The discussion will also prove to provide a useful starting point for the systematic construction of basis sets used in the more complicated many-electron calculations reported below.
Let $E_0^{(N)}$ be the ground state energy of a $N$-electron QD.
Let us estimate  variationally the ground state energy of one electron in a Gaussian potential.
Taking into account that, around its center, the Gaussian potential resembles a parabolic one, we propose a a normalized $s$-type Gaussian trial function of exponent $\alpha$
\begin{equation}
\varphi_s(r)=(2\alpha/\pi)^{3/4}\exp(-\alpha r^2).
\end{equation} 
The deeper the well, the more similar its ground state is to that of the harmonic oscillator.
The expectation value of the one-electron Hamiltonian then becomes
\begin{equation}
\label{E_alpha}
E(\alpha) = 3\alpha - V_0 \left(\frac{2\alpha}{\lambda+2\alpha}\right)^{3/2},
\end{equation}
where the first and second terms are the mean values of the kinetic energy and the confining potential, respectively. Minimization with respect to $\alpha$, {\em i.e.}  $\partial E(\alpha)/\partial\alpha=0$, gives
\begin{equation}
\label{eq for alpha}
(\lambda+2\alpha)^5 - 8\lambda^2V_0^2\alpha =0,
\end{equation}
which has to be solved numerically for given $\lambda$ and $V_0$. Eq. (\ref{eq for alpha}) has five roots but, typically, the one that minimizes $E(\alpha)$, is a positive number of the order of unit $\alpha=\alpha_{\rm opt}$. 
The other roots can be rejected on a physical basis because whether they are complex conjugate pairs, or real but negative or very close to zero and, hence,  the corresponding trial functions do not describe bound states. Thus, within this variational approach, the energy of a single-electron QD is
\begin{equation}
E_0^{(1)} = E(\alpha_{\rm opt}).
\end{equation}
On the other hand, the ground state of a two-electron QD corresponds to one electron with spin $+\frac{1}{2}$ and the other with spin $-\frac{1}{2}$ and interacting with each other through the Coulomb interaction $V_{ee}=2/r_{12}$. Therefore,
%Then, within the UHF approximation,
%\begin{equation}
% \varepsilon_0=E^{(1)}_0 + J,
% \end{equation}
%and 
%\begin{equation}
%E^{(2)}=2\varepsilon_0-J=2E^{(1)}_0+J.
%\end{equation}
\begin{equation}
\label{2E_1+J}
E^{(2)}=2E^{(1)}_0+J,
\end{equation}
where $J$ is the energy shift of the one-electron levels due to the Coulomb interaction between two Gaussian charge densities
\begin{equation}
\label{J value}
J=2\int |\varphi_s({\bf r}_1)|^2 \frac{1}{r_{12}}|\varphi_s({\bf r}_2)|^2 d^3{\bf r}_1 d^3{\bf r}_2
= 4\sqrt{\frac{\alpha_{\rm opt}}{\pi}}.
\end{equation}
The condition of stability of the two-electron QD discussed in Refs. \cite{Szafran99,Bednarek99} and applied in Refs. \cite{Adamowski00, Boyacioglu07}, $E_0^{(2)}<E_0^{(1)}$, is equivalent to
\begin{equation}
E_0^{(1)} < -J. 
\end{equation}
We shall discuss thoroughly the conditions for stability below when considering the results for few-electron QDs.\\

In the calculations reported in this work we have considered both one- and few-electron systems. The former were used for benchmark purposes and provide a systematic way for calibrating the method. A one-electron Gaussian potential of $V_0=400 R_D$ and $\lambda=1$ has been considered by various authors for the comparison of the methods proposed in the literature. We also address such a case as our first calculation to assess the accuracy of our methodology. The first issue to be considered is the choice of a suitable Cartesian Gaussian basis set $\{\varphi_\ell^{(i)} \}$, $(i=1,\ldots, K)$, where 
\begin{equation}
\varphi_\ell^{(i)} = x^m y^n z^p \exp(-\alpha_i r^2),
\end{equation}
and $\ell = m+n+p$ is the angular momentum of the function. Hereafter we use the spectroscopic notation $s$, $p$, $d,\ldots$ for $\ell=0, 1, 2,\ldots$, thus
\begin{equation}
\{\varphi_\ell^{(i)} \}= \{\varphi^{(i)}_s,\varphi^{(i)}_{p_x},\varphi^{(i)}_{p_y},\varphi^{(i)}_{p_z},\varphi^{(i)}_{d_{xx}},\varphi^{(i)}_{d_{xy}},\ldots\}.
\end{equation}

The one-electron QDs were solved by direct diagonalization of the Hamiltonian, while the many-electron systems were treated with the unrestricted Hartree-Fock (UHF) method. The calculations for a given potential, {\em i.e.}, for a given pair ($V_0$,$\lambda$), were performed with the same basis sets, irrespective of the number of electrons.\\

 The prescription for the choice of the basis is based on the previously discussed variational energy expression [Eq. (\ref{E_alpha})], and the optimal exponent $\alpha_{\rm opt}$ [Eq. (\ref{eq for alpha})].
%===========================================================================

The values of $\alpha_{\rm opt}$ for the three cases considered in this paper are listed in Table \ref{potenciales}. Those exponents were then used to generate the basis functions with higher angular momentum, {\em i.e.}, we choose $\alpha_s=\alpha_p=\alpha_d=\ldots=\alpha_{\rm opt}$. This procedure can be justified 
as long as the Gaussian potential is similar to the parabolic potential, whose solutions are Hermite polynomials multiplied by Gaussian functions with an exponent independent of the principal and angular quantum numbers $n$ and $\ell$, thus exactly satisfying the prescription given above.
By including one Gaussian function with the optimized exponent in each angular momentum block up to a maximum value $L$, {\em i.e.} $0\le\ell\le L$, the low-lying states are reasonably well reproduced as compared to previously reported values. In particular, the ground state energy is correct within $10^{-2}R_D$ approximately, as expected, because of the variational procedure for obtaining the basis exponents. Generally, states having $\ell >L$ are missing or, if obtained, have the larger errors.
This is also a consequence of the fact that the low-lying states of the parabolic and Gaussian potentials are alike, but those of high angular momentum are not.
Hence, to get a correct description of high-lying excited states, functions with the corresponding angular momentum $\ell$ have to be included. In atomic and molecular calculations, a useful procedure for enlarging the basis set has been the so-called {\em even-tempered} criterion  \cite{Davidson86} by which a basis set having various functions of the same angular momentum has their exponent in the same ratio, {\em i.e.}, for a given $\ell$, the exponents $\alpha^{(1)}_\ell, \alpha^{(2)}_\ell, \alpha^{(3)}_\ell,\ldots$ are in the ratio $\alpha^{(i+1)}_\ell/\alpha^{(i)}_\ell = {\rm const}$. 
In the limit of a large number of functions the basis set should become complete, independently of the ratio chosen. We take arbitrarily a ratio of 2 to enlarge every block of angular momentum in the original basis set, thus including exponents $\alpha_\ell^{(i)}=\alpha_{\rm opt}/2^{i-1}$ $(i=1,\ldots,K)$ smaller than the optimized one. A value of $K=4$ has been enough to reach converged energies within $10^{-3}R_D$ for all the calculations reported here.
In summary, the basis set used in all calculations consists of a set $\{\varphi_s^{(i)}, \varphi_p^{(i)}, \ldots, \varphi_\ell^{(i)},\ldots\}$, with $i=1,\ldots,K=4$ and $\ell\le L=4$, having 140 Cartesian Gaussian functions.\\
%===================================================================================
\begin{table}
\caption{\label{potenciales} Values of the parameters $\lambda$ and $V_0$ for the Gaussian potentials studied. The radii of the quantum dots are given by $R=1/\sqrt{2\lambda}$. $\alpha_{\rm opt}$ are the variationally optimized exponents used to construct the basis sets.}
\begin{center}
\begin{tabular}{cccc}
\hline\hline
 $\lambda$ (a.u.$^*$) & $V_0$ ($R_D$) & $R$ ($a_D$) & $\alpha_{\rm opt} (a_D^{-2})$ \\ \hline
      1.0             &  400            &      0.707    & 9.370 885 \\
      0.5             &   50            &      1.0      & 2.183 148 \\
      0.5             &   15            &      1.0      & 1.047 992 \\
\hline\hline
\end{tabular}
\end{center}
\end{table}
%%%%%%%%%%%%%%%%%%%%%%%%%%%%%%%%%%%%%%%%%%%%%%%%%%%%%%%%%%%%
\section{Results and discussion}\label{resul}
%===================================================================================
\subsection{One-electron quantum dots}
%%%%%%%%%%%%%%%%%%%%%%%%%%%%%%%%%%%%%%%%%%%%%%%%%%%%%%%%%%%%%%%%%%%%%%%%%%%%%%%%%%
\begin{table}
\caption{
\label{one-electron}
Bound states energies of the one-electron spherical Gaussian potentials  $(V_0,\lambda)=(400,1),\ (50,0.5)$ and $(15, 0.5)$ calculated by diagonalization of the Hamiltonian matrix and Numerov's integration method. Energies are given in donor Rydberg units $R_D$ and lenths in donor Bohr radius $a_D$, and dots stand for positive-energy states. }
\begin{tabular}{cccccccccc}\hline\hline
$n$\ $\ell$ & \multicolumn{3}{c}{$V_0$=400 R$_D$} &&  \multicolumn{2}{c}{$V_0$=50 R$_D$}  && \multicolumn{2}{c}{$V_0$=15 R$_D$}  \\
\cline{2-4}\cline{6-7} \cline{9-10}
       & Diagon.  & Numerov   & Ref.\cite{Lai83} && Diagon. & Numerov && Diagon. & Numerov \\
\hline
 1$s$  &    -341.895     & -341.892 & -341.8952&& -35.958  & -35.958  && -7.762  &   -7.762  \\
 1$p$  &    -304.463     & -304.463 & -304.4628&& -27.282  & -27.282  && -3.697  &   -3.698    \\
 2$s$  &    -269.644    &  -269.640 & -269.6445&& -19.987  & -19.987  && -1.215  &   -1.215  \\
 1$d$  &    -268.110     & -268.111 & -268.1107&& -19.204  & -19.204  && -0.439  &   -0.439    \\
 2$p$  &    -235.446     & -235.450 & -235.4500&& -13.109  & -13.111  && $\ldots$ &  $\ldots$ \\
 1$f$  &    -232.849     & -232.875 & -232.8753&& -11.780  & -11.785  && $\ldots$ &  $\ldots$  \\
 3$s$  &    -203.983    &  -203.979 & -203.9835&&  -7.800  &  -7.800  && $\ldots$ &  $\ldots$  \\
 2$d$  &    -202.427     & -202.431 & -202.4313&&  -7.009  &  -7.010  && $\ldots$ &  $\ldots$  \\
 1$g$  &    -198.700     & -198.798 & -198.7983&&  -5.111  &  -5.122  && $\ldots$ &  $\ldots$  \\
 3$p$  &    -173.156     & -173.244 & -173.2443&&  -3.173  &  -3.179  && $\ldots$ &  $\ldots$  \\
 2$f$  &    -167.797     & -170.639 & -170.6393&&  -1.864  &  -1.876  && $\ldots$ &  $\ldots$  \\
 4$s$  &    -145.372     & -145.373 & -145.3779&&  -0.598  &  -0.600  && $\ldots$ &  $\ldots$  \\
 3$d$  &    -143.741     & -143.809 & -143.8091&& $\ldots$ &  $\ldots$ &&$\ldots$ &  $\ldots$ \\
 2$g$  &    -138.045     & -140.135 & -140.1351&& $\ldots$ &  $\ldots$ &&$\ldots$ &  $\ldots$ \\
\hline\hline\\
\end{tabular}
\end{table}
%%%%%%%%%%%%%%%%%%%%%%%%%%%%%%%%%%%%%%%%%%%%%%%%%%%%%%%%%%%%%%%%%%%%%%%
In Table \ref{one-electron} we have listed the energy eigenvalues calculated for the potentials studied. The first one, $V_0=400$, has been studied previously \cite{Bessis82,Lai83, Cohen84, Chatterjee85} and provides a measure of the precision of our calculations. The first column refers to energies calculated by diagonalizing the matrix representation of the Hamiltonian after expansion in the basis set described above. The second column corresponds to a numerical solution of the Schr\"odinger equation by using the Numerov's integration algorithm \cite{Thijssen}, that we have implemented as an independent verification for the basis set expansion method. For the first potential, where results from a number of methods are available, the third column referes to Ref. \cite{Lai83}, which has probably the most accurate results reported so far. This potential has bound states with angular momenta as high as $\ell=7$, what makes it difficult to reproduce with a basis set having functions with $\ell=0$ through $\ell=4$. This is particularly apparent for the  energies $E_{2f}$ and $E_{2g}$, having the largest errors. This effect of the incompleteness of the basis set, can be removed by adding four extra functions for each $\ell$ value and symmetry with $5\le\ell\le 7$, what amounts a total of 360 basis functions. The energies calculated with this enlarged basis become $E_{2f}=-170.639R_D$ and $E_{2g}=-140.133R_D$, in good agreement with Ref. \cite{Lai83}. Nevertheless, the need of such a more demanding calculation should be evaluated taking into account whether those states are to be occupied or because there is a perturbation exciting the electrons from low-lying states to them.

The case $V_0=50 R_D$ shows a good agreement between the direct diagonalization and numerical integration, implying that both the methodology and the basis set used are accurate enough. The scheme of levels obtained are in agreement with Ref. \cite{Adamowski00}. The system $V_0=15 R_D$ has not been treated previously and it provides a system having just a few bound states, in which the parabolic approximation could be poor.
 
It also can be seen from Table \ref{one-electron} that the energy of the corresponding parabolic potential, $\hbar\omega=\sqrt{2\lambda V_0/m}=5.48,\ 10,$ and 40 $R_D$, for $V_0=15$, 50 and 400 $R_D$ respectively, represent approximately, the energy difference between the ground and first-excited state. If the potential were really parabolic, that energy difference would be the same for every pair of consecutive states. That is not the case for the Gaussian potential, {\em i.e.}, the higher the pair of states considered lie, the worse the approximation becomes, as already noted by other authors \cite{Adamowski00, Boyacioglu07}. The basis sets optimized for these one-electron calculations were hereafter used for the few-electrons systems.
\subsection{Few-electron quantum dots}
%###################################################################################

%%%%%% Sergio

\begin{figure}
\caption{\label{fig1} 
Ground state energies $E_0^{(N)}$ of $N$-electron Gaussian quantum dots ($N\le 10$) as a function of the dot radii, for depths $V_0=15 R_D$ (left) and $V_0=50 R_D$ (right). The cross-over of levels occurrs when $E_0^{(N+1)}\le E_0^{(N)}$ at a critical radii $R_c^{(N)}$, starting from approximately $R_{c}^{(1)}=0.43 a_D$ and $R_{c}^{(1)}=0.22 a_D$, respectively.}
\begin{tabular}{cc}
\includegraphics[scale=0.5]{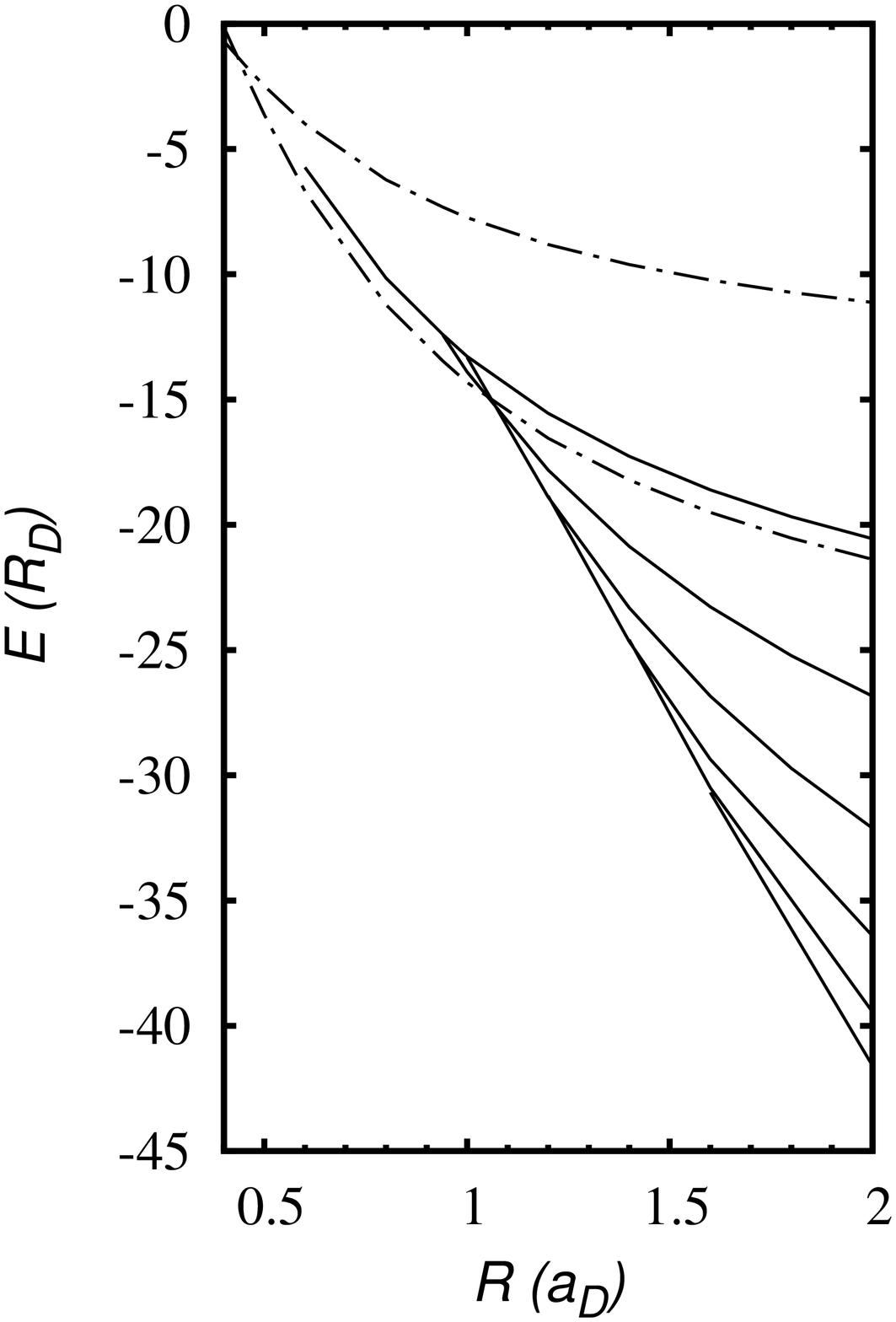} &
\includegraphics[scale=0.5]{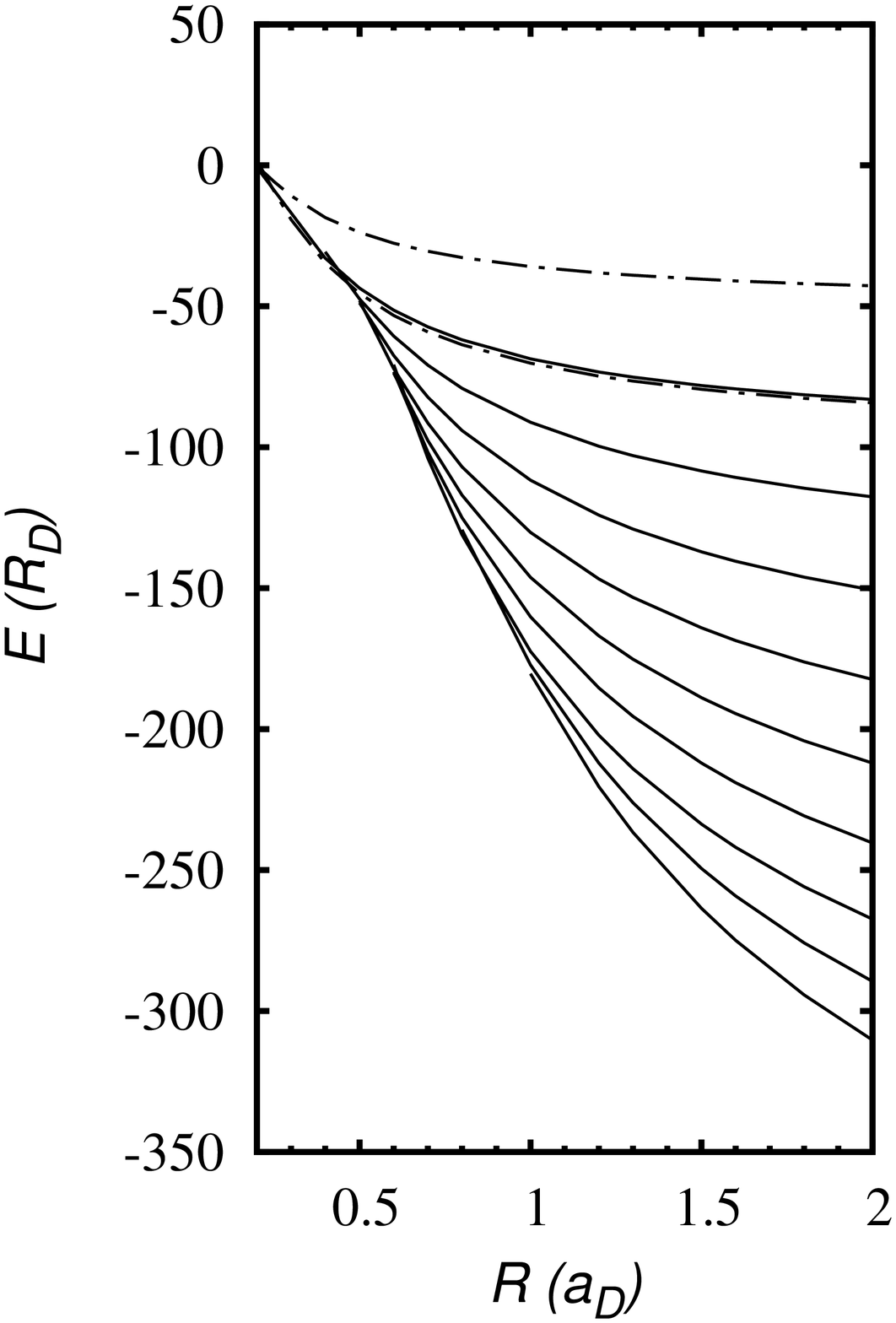} 
\end{tabular}
\end{figure}

The results of Hartree-Fock calculations of the ground state energy of $N$-electron systems ($N=1,\ldots,10$), as a function of the QD radius, for $V_0=15 R_D$ (left panel) and $V_0=50 R_D$ (right panel), are shown in Fig. \ref{fig1}. 
This figure has features alike the one obtained with a finite depth square well \cite{Bednarek99}. 
For comparison purposes, the variational one- and two-electron energies $E_0^{(1)}$ [Eq. (\ref{E_alpha})] and $E_0^{(2)}$ [Eq. (\ref{2E_1+J})] were also plotted (dashed lines).

The figure shows several critical radii $R_{c}^{(N)}$ at which a crossover between the $E_0^{(N)}$ and $E_0^{(N-1)}$ ground states occurs. Thus, $R_{c}^{(N)}$ is the minimum radius to have $N$ bound electrons; for $R>R_{c}^{(N)}$, the $N$-electron QD becomes more stable than the $(N-1)$-electron one. It should be noted that at small enough radii no bound state exists. At radii less than $R_{c}^{(2)}=0.43 a_D$ (for $V_0=15 R_D$) and $R_{c}^{(2)}=0.23 a_D$ (for $R=50 R_D$) only a single electron can be bound, but for $R>R_{c}^{(2)}$,  $E_0^{(2)}<E_0^{(1)}$. The radii for $N=1$ and $N=2$ are quite similar ($R_{c}^{(1)}\simeq R_{c}^{(2)}$). So also happens for $R_{c}^{(3)}\simeq R_{c}^{(4)} \ldots\simeq R_{c}^{(8)}$ and $R_{c}^{(9)}\simeq R_{c}^{(10)}$.

Most of the general features of the stability of the $N$-electron system can be understood considering the one-electron energies. 
An electron confined within a well of a typical size $R$ will have a momentum of order $\hbar/R$ due to the uncertainty principle; its kinetic energy then becomes of order $\hbar^2/2mR^2$, which approaches zero as $R$ goes to infinity. On the other hand, its potential energy $-V_0 \exp(-r^2/2R^2)$ approaches the bottom of the well $-V_0$. Then, in the limit of large radius, the energy of the one-electron system goes to $-V_0$ and, since also the Coulomb interaction goes to zero, the energy of the $N$-electron well approaches $-NV_0$. Such a trend can be observed in Fig. 1.

At a given finite radius, however, the energies $E_0^{(N)}$ are not equally spaced for successive $N$, because of the shell structure of the energy levels of the potential well. This shell structure is even more strikingly revealed in the left panel of Figure 2, where the chemical potential $\mu(N)$ is depicted as a function of the QD radius. The chemical potential represents the affinity of the well for binding an extra electron. It is the equivalent of the ionization potential or the electron affinity in atomic physics. It is clearly apparent a grouping of the lines corresponding to the number of electrons occupying the same one-electron level. All curves decreases as $R$ increases due to the fact that when the electrons becomes less confined, the system approaches a classical behavior and it is easier to add a new electron.
In the right panel of Figure 2, the charging energy, defined as the difference between the chemical potential of two systems differing in one electron, {\em i.e.}  $E_{char}^{(N)} =\mu(N+1)-\mu(N)$, is depicted as a function of the number of electrons in the QD for three potential radii, namely, 0.8, 1.5 and 2 $a_D$. The charging energy gives a measure of the stability of the system, the larger $E_{char}$, the more stable the system. For $N=2, 5$ and 8 electrons, the system presents large values of $E_{char}$, in correspondence with the number of electrons needed for filling or half-filling a shell. The peaks heights also diminishes as the QD radius increases. That is, the charging energy as a function of $N$, tends to be flat because of the disappearance of the shell structure in this classical limit.

As an illustration of the relation between the QD depth and the critical radius, we have depicted, in Figure \ref{phase_diagram}, the curves $V_0$ versus $1/R_c^2$ for a QD of three interacting (upper curve) and non interacting electrons (lower curve). They are the locus of the points representing the minimum QD radii for a given depth. Both curves are nearly straight lines, thus showing that $V_0 R_c^2$ is approximately constant. 
We shall show, in the following, that the variational energy of the one-electron Gaussian potential also has such a property. By defining the dimensionless variable $x=2\alpha/\lambda$, the energy $E_0(\alpha)$ [Eq. (\ref{E_alpha})] can be written as
\begin{equation}
E(\alpha) = \lambda \left[\frac{3}{2}x-2C\left(\frac{x}{x+1}\right)^{3/2}\right]=\lambda\varepsilon(C),
\end{equation}
where $C=V_0R^2$ and $\varepsilon$ is the dimensionless energy depending on $C$. 
On the other hand, the equation determining the optimal variational exponent can be written as
\begin{equation}
\label{opt exp adimens}
\frac{\partial\varepsilon}{\partial x} = \frac{3}{2}- 3C\frac{x^{1/2}}{(x+1)^{5/2}}=0.
\end{equation}
The equation $\varepsilon(C)=0$ represents the condition to have at least one bound state, which clearly only depends on $V_0R^2$.
This equation and (\ref{opt exp adimens}) are simultaneously satisfied if $x=1/2$ and $C=9\sqrt{3}/8$, {\em i.e.}, $\alpha=\lambda/4=1/8R^2$ and $V_0R^2=1.95$. Hence, for a given $V_0$, the minimal radius for having one bound electron is $R_c^{(1)}=\sqrt{1.95/V_0}=1.40/\sqrt{V_0}$.

This can be compared to the results we would obtain with the truncated parabolic approximation of the Gaussian potential,
{\em i.e.}, $V(r)=-V_0+m\omega^2r^2/2$ if $r<R$, and zero otherwise, with $\omega$ given by Eq. (\ref{omega}). In such a case, the condition for having at least one bound state is that the zero point energy becomes less or equal than zero. That is, at the critical radius,

\begin{equation}
E_0^{(1)}= -V_0+\frac{3}{2}\hbar\omega = -V_0+\frac{3}{2}\hbar\sqrt{\frac{V_0}{mR^2}}=0,
\end{equation}
thus giving $V_0R_c^2=9/4=2.25$, {\em i.e.}, $R_c^{(1)}=1.5/\sqrt{V_0}$, comparable to the relation obtained above. If there were no interactions between the electrons, the energy for allocating the second electron would be the same, because both of them would occupy the one-electron ground state, although with opposite spins. The Coulomb interaction between the electron pair, however, modifies such an energy by an amount $J$. Nevertheless, $J$ is, for small- and medium-size QDs, much smaller than the energy difference between the ground and first excited one-electron states. Thus, the next critical radius $R_c^{(2)}$ is near to the first one $R_c^{(1)}$. On the other hand, the radius $R_c^{(3)}$ includes the effects of both the electron-electron interaction and the shell structure because the third electron has to occupy the first excited one-electron level. We can estimate this radius by using an argument similar to the one above. The first excited state of the harmonic oscillator is now $(5/2)\hbar\omega$ above the bottom of the well, thus giving $V_0R_c^2=25/4=6.25$.
A fit of the plot of Figure \ref{phase_diagram} of the form $V_0=a+b R_c^\gamma$ gives $V_0=6.30 R_c^{-2}$ for the non interacting case, which compares fairly well to the estimation from the truncated harmonic oscillator, thus showing that the non-interacting picture is qualitatively correct. Nevertheless, the corresponding fit of the UHF calculations gives a relation $V_0=7.4+9.0 R^{-1.88}$ as a consequence of the Coulomb and exchange interactions.\\

Furthermore, the fact that $V_0R^2$ determines the number of bound electrons is a property shared by a wealth of types of potentials.
Let $H$ be a Hamiltonian of the form
\begin{equation}
H=-\nabla^2 -V_0 u\left(\rho\right) 
\end{equation}
where the potential can be factored as a product of a typical energy $-V_0$ times a function $u(\rho)$ of the dimensionless variable $\rho=r/R$.
That includes most of the types of confining potential used in previous works, {\em e.g.}, the square well $V(r)=-V_0\theta(1-\rho)$, the parabolic potential $V(r)=\hbar\omega (r/R)^2$ with $R=\sqrt{\hbar/2m\omega}$ and, clearly, the Gaussian potential. Expressing the Laplacian in terms of $\rho$, we get
\begin{equation}
H=\frac{1}{R^2}\left[-\nabla_{\rho}^2 -V_0R^2 u(\rho) \right],
\end{equation}
which has energy eigenvalues $\varepsilon_n (V_0,R)= \epsilon_n(V_0R^2)/R^{2}$.
Neglecting the electron-electron interaction, the condition of stability for a $N$-electron system is just $\varepsilon_n<0$, where $n$ is the quantum number corresponding to the highest occupied state. It is fulfilled for certain values of $V_0R_c^2=C_i$, giving rise to stability regions.

Therefore, when the electron-electron interaction $V_{ee}$ can be neglected, the plot of $V_0$ versus $1/R_c^{2}$ represents a phase diagram constituted by regions delimited by straight lines delimiting the various zones where $N$ electrons can form a bound state. The argument does not hold when $V_{ee}$ is not negligible, since it scales with $R$ differently to the kinetic energy term.\\
%%%%%%%%%%%%%%%%

It is interesting to analyze how the electrons occupy the UHF spin-orbitals as  they are added into the system.
Table \ref{config} shows the most stable UHF configurations, for given $N$, along with the total spin projection $M_S$, for $V_0=50 R_D$ and $R \ge R_c^{(10)}$.
In all the calculations performed, with $N\le 10$, Hund rule is fulfilled, {\em i.e.}, electrons in the same shell maximize the projection of the total electronic spin, in agreement with previous works. 
%===================================================================================
\begin{table}
\caption{\label{config} Electronic configuration and $z$-projection of total spin $M_S$ for the most stable UHF configurations of a $N$-electron Gaussian potential of $V_0=50R_D$, calculated in the present work.
The spin projections of the orbital are represented by $\uparrow$ and $\downarrow$, for +1/2 and -1/2, respectively. Spin indices are omitted for closed shell configurations. Hund Rule is fulfilled for all the configurations shown.}
\begin{center}
\begin{tabular}{cccc}
\hline\hline
 $N$        & $M_S$   & Electronic Configuration  \\ \hline
      1     &  1/2 & $1s_{\uparrow}$  \\
      2     &  0   & $1s^2$ \\
      3     &  1/2 & $1s^2 1p_{\uparrow}$ \\
      4     &  1   & $1s^2 1p_{\uparrow}^2$ \\
      5     &  3/2 & $1s^2 1p_{{\uparrow}}^3$ \\
      6     &  3/2 & $1s^2 1p_{{\uparrow}}^3 1p_{{\downarrow}}^1$ \\
      7     &  1/2 & $1s^2 1p_{{\uparrow}}^3 1p_{{\downarrow}}^2$ \\
      8     &  0   & $1s^2 1p^6$ \\
      9     &  1/2 & $1s^2 1p^6 2s_{\uparrow}$ \\
      10     &  0  & $1s^2 1p^62s^2$ \\      
\hline\hline
\end{tabular}
\end{center}
\end{table}
%%%%%%%%%%%%%%%%%%%%%%%%%%%%%%%%%%%%%%%%%%%%%%%%%%%%%%%%%%%%
As an example, we consider the UHF levels of a $V_0=15R_D$ well of $R=1 a_D$.
Figure \ref{levels} shows how the scheme of levels changes as the first three electrons are added to it.
Panel (b) shows the one-electron levels of the potential with a single electron occupying the ground state $1s$. When a second electron is added to the system, the pair can form a state with total spin projection $M_S=1$ [panel (a)] or $M_S=0$ [panel (c)]. 
Panels (a) and (d) shows the states available for spins $+\frac{1}{2}$ and $-\frac{1}{2}$ drawn adjacent to each other.
The electron-electron repulsion shifts and splits the one-electron levels through the Coulomb and exchange interactions. 
In the $M_S=0$ configuration, both electrons can occupy the same orbital, while the $M_S=1$ configuration requires the second electron to occupy the higher $p_\uparrow$ orbital, thus giving a higher total energy. The effect of adding a third electron to the $M_S=0$ configuration, is shown in panel (d). The $1s$ orbital shifts upwards and splits into $1s_\uparrow$ and $1s_\downarrow$, while the $1p_\uparrow$ becomes stabilized.

%=====================================================

%%%%%%%%%%%%%%%%%%%%%%%%%%%%%%%%%%%%%%%%%%%%%
\begin{figure}
\caption{
Chemical potential (left) and charging energy (right) of the $N$-electron Gaussian potentials considered in this work. The grouping of lines in the chemical potential $\mu$ and the peaks in the charging energy show the shell structure.}
\begin{tabular}{cc}
\includegraphics[scale=0.5]{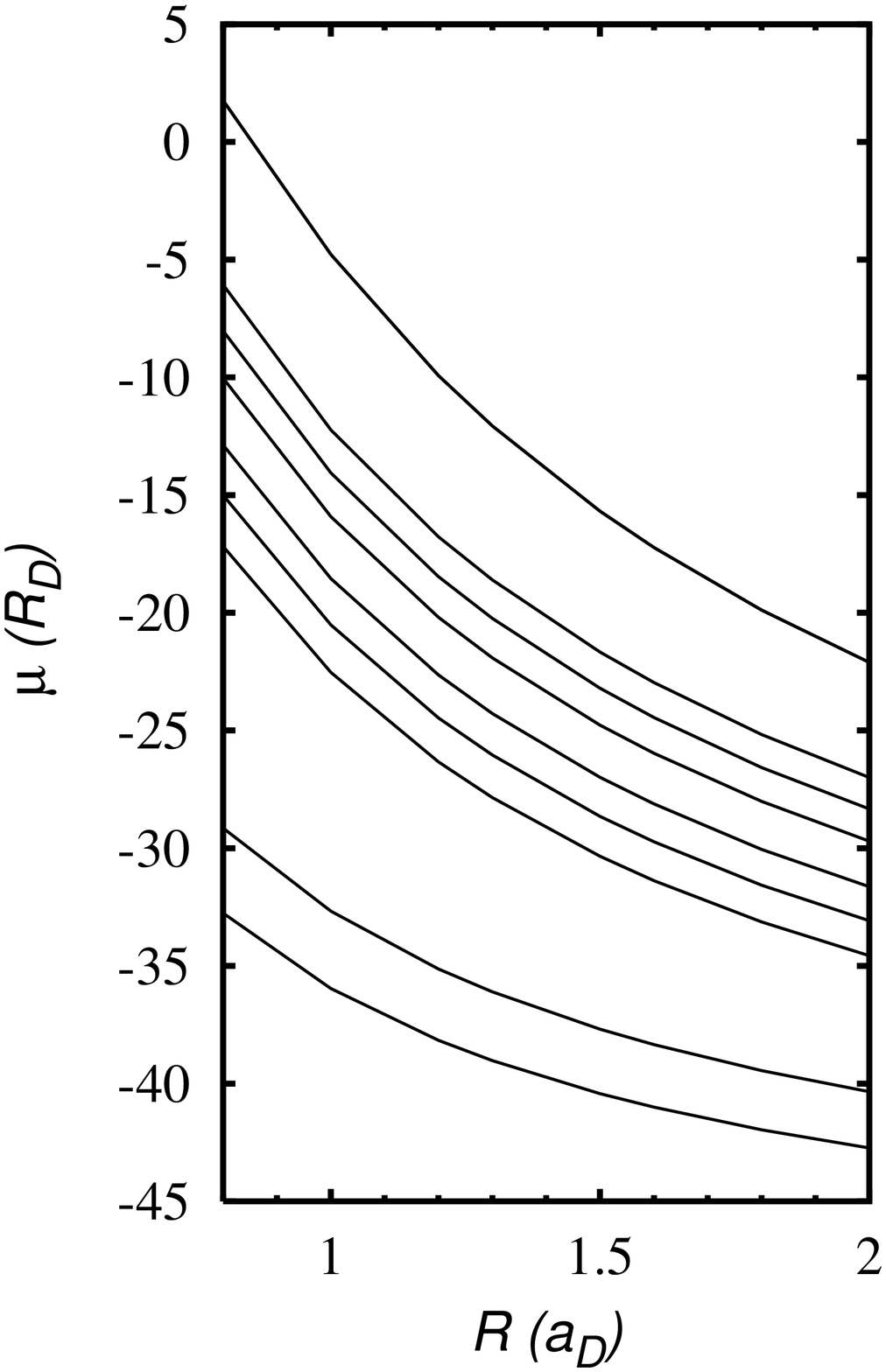} &
\includegraphics[scale=0.5]{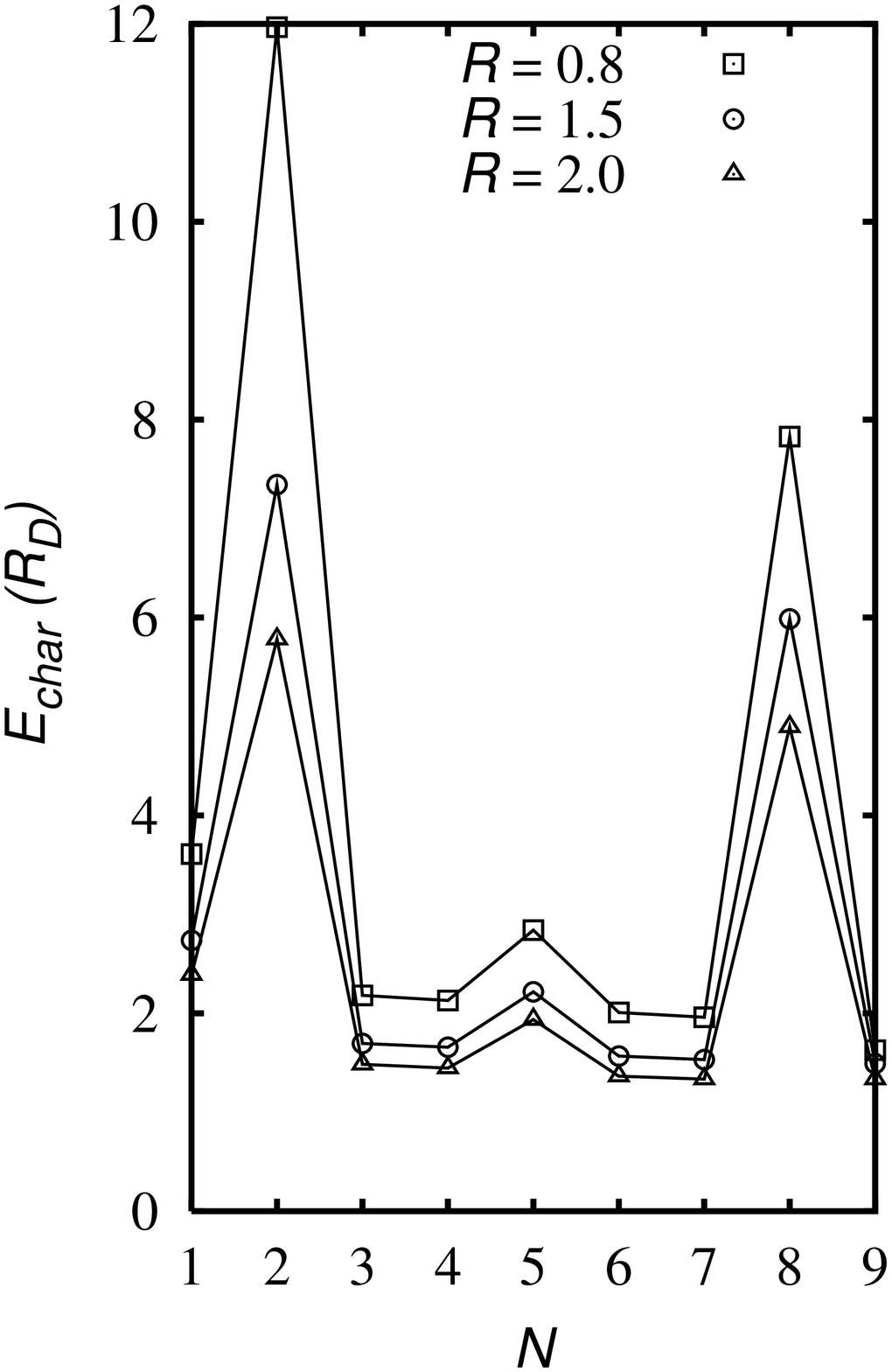}
\end{tabular}
\end{figure}
%%%%%%%%%%%%%%%%%%%%%%%%%%%%%%%%%%%%%%%%%%%%%
\begin{figure}
\caption{
Scheme of the UHF levels for the $V_0=15R_D$ and $R=1a_D$ Gaussian well having one two and three bound electrons. $\uparrow$ and $\downarrow$ refers to the spin projections. In the cases $N=2$, $M_S=1$ [panel (a)] and $N=3$, $M_S=1/2$ [panel (d)], the states for spins $\uparrow$ and $\downarrow$ are drawn separately.
\label{levels}}
\begin{center}
\includegraphics[scale=0.5]{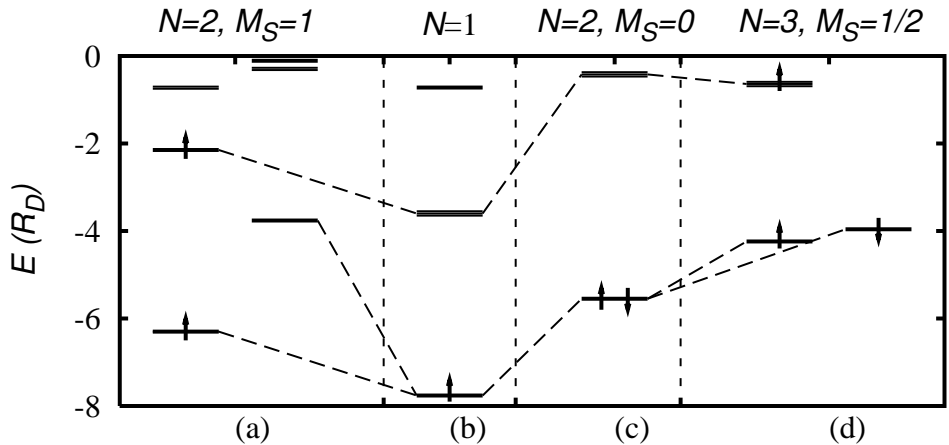} 
\end{center}
\end{figure}
%%%%%%%%%%%%%%%%%%%%%%%%%%%%%%%%%%%%%%%%%%%%%
\begin{figure}
\caption{
Critical lines representing the relation between the potential depth $V_0$ and $1/[R^{(3)}_c]^2$, where $R_c^{(3)}$ is the minimal radius to have three bound electrons in the system. The lower curve represents the critical line for the non interacting problem and its fit to the straight line $V_0=6.3 R^{-2}$. The upper curve represents the critical UHF line. The dashed line corresponds to the fit $V_0=7.4+9.0 R^{-1.88}$.
 \label{phase_diagram}}
\includegraphics[scale=0.5]{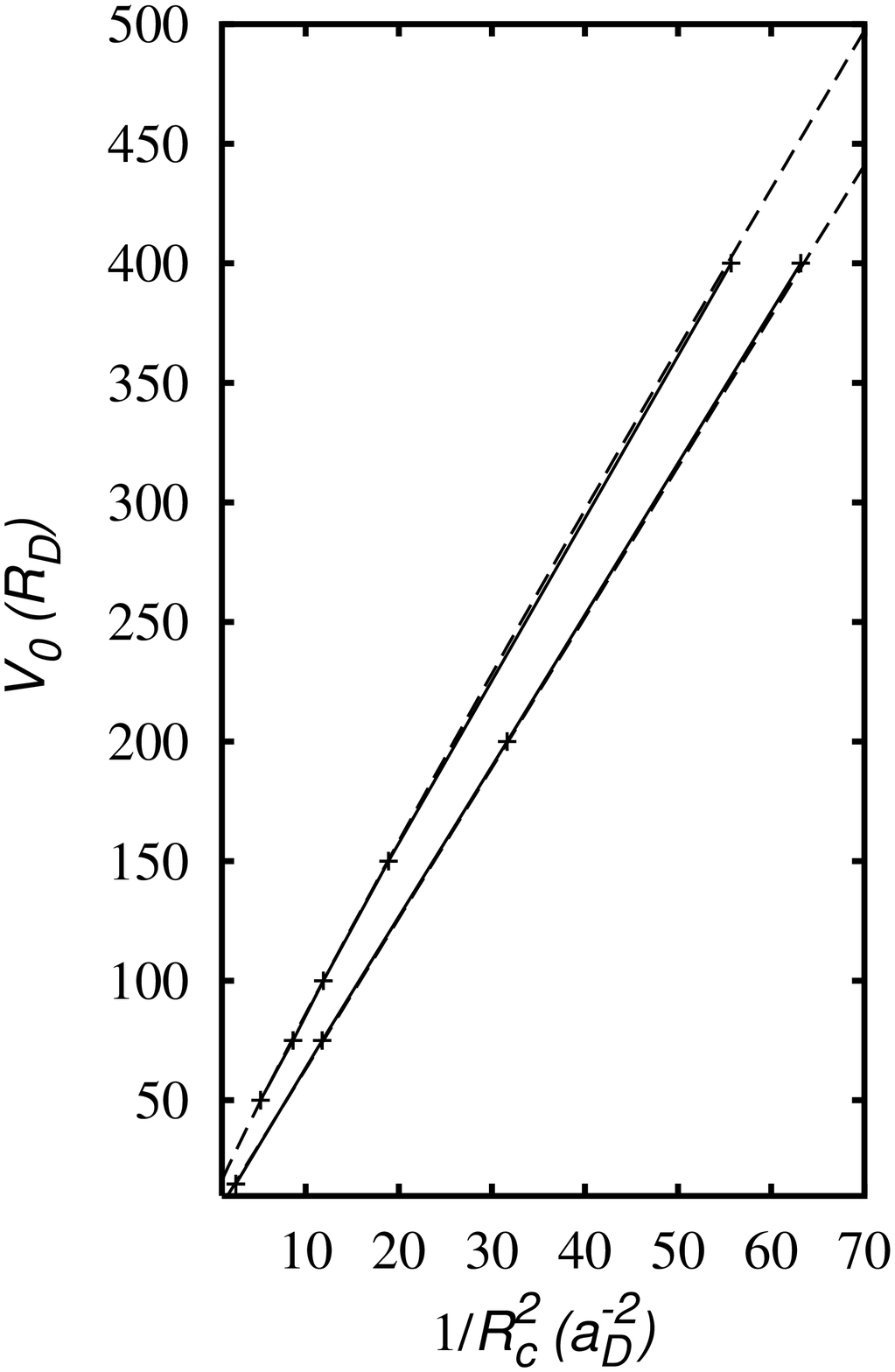} 
\end{figure}
%%%%%%%%%%%%%%%%%%%%%%%%%%%%%%%%%%%%%%%%%%%%%
\section{Conclusions}
In summary, we have calculated the electronic structure of electrons confined in a Gaussian QD by using the UHF method in its Pople and Nesbet's form. 
We gave a systematic procedure for constructing the basis set for the self-consistent calculations from the one-electron problem and with a given precision. 
The orbital energies provide an insight of the occupancy of the levels and show the fulfillment of Hund rule in agreement with previous results obtained with other models \cite{Bednarek99, Destefani04}.

Since the Gaussian potential has a finite depth, binding and dissociation processes can occur. The criterion for the stability of a $N$-electron system was already discussed in Ref. \cite{Szafran99} in terms of its energy $E_0^{(N)}$ as compared to $E_0^{(N-1)}$. The condition $E_0^{(N)} < E_0^{(N-1)}$ is equivalent, in UHF calculations, to the condition that the highest-occupied molecular orbital (HOMO) is bound, {\em i.e.}, $\varepsilon_{\rm HOMO}<0$, as a consequence of the Koopman's theorem \cite{Szabo}. If the electron-electron interaction can be neglected, the regions of the $(V_0,R)$ plane, having $N$ bound electrons, are delimited by the relation $V_0R^2={\rm const}$. This feature has been proven to be a widely valid property, as a consequence of the scaling properties of the confining potential.  

Both the chemical potential and the charging energy show a shell structure with peaks of stability at $N=2$, 5 and 8, corresponding to filled and half-filled shells. Both magnitudes decreases as the confinement range increases.
The methodology applied in this work to single QDs, can straightforwardly be transferred to the study of the electronic structure and properties of QDs arrays. It is also amenable for considering correlation effects with standard atomic methods. Currently, work is in progress along such lines.
\section*{Acknowledgements}
This work has been supported by Argentine National Council for Science and Technique CONICET and SGCyT (Universidad Nacional del Nordeste).
%##################################################################################

%#################################################################################
\end{document}